# Static Analysis Deployment Pitfalls

Flash Sheridan

Code Integrity Solutions
Palo Alto, California USA
flash@pobox.com   http://pobox.com/~flash

*Abstract*—Organizational, political, and configuration mistakes in the deployment of a static source code analysis tool within a software development organization can result in most of the value of the tool being lost, even while apparently meeting management goals. A list of pitfalls encountered as a static analysis consultant is presented, with discussion of techniques for avoiding or mitigating them. This is part of a work in progress, tentatively entitled "Handbook of Static Analysis Deployment."

## I. INTRODUCTION

In the last decade, a revolution in static source code analysis tools enabled their widespread deployment for bug-finding in software development organizations. But such tools can lose most of their value if organizational, political, and configuration issues are not handled skillfully. The natural approach may seem to management to be succeeding, while in the long run being nearly useless, and in the meantime bringing static analysis into disrepute with the programmers dealing with the supposed defects. I present here some of the leading pitfalls I have encountered in deploying, maintaining, and using static analysis tools within software development organizations, principally Coverity at Palm, Inc. and Klocwork as a consultant for Code Integrity Solutions.

## II. PITFALL 1: DON'T LOSE FOCUS ON FINDING IMPORTANT BUGS

The primary goal of static analysis is, or at least ought to be, the improvement of code by finding and fixing defects. But finding bugs is a messy process, more familiar to software testers than to those who are usually given responsibility for static analysis: software developers or tool maintainers [Yang & Sheridan 2010, p. 16]. The imperatives for the latter two groups are generally a clearly-defined process and a smoothly running tool. These imperatives conflict with the rapid changes in configuration needed to adapt to experience in finding bugs. For decisions about finding bugs, the final say must belong to the software quality group, not development or tool maintenance [Spolsky 2004, chapter 22].

The conventional wisdom recommends a management champion to provide political muscle behind static analysis, to keep the focus on fixing bugs, and to overcome the tendency to deprioritize bugs found by the tool [Bessey, Engler, et al. 2010]. Ship-at-All-Costs-itis is pervasive in the software industry: Quality generally has fewer constituents than the schedule.

## III. PITFALL 2: DON'T BUY A TOOL BASED ON BUGS IT FINDS IN OTHER PEOPLE'S CODE

Before you commit to a static analysis tool, make sure that it finds important bugs in your real code. Bugs found in open source or demo code can be very impressive; but your organization's code, while it's under development (which is the cheapest time to find bugs) will be very different from code which has already been made public. Also, getting a static analysis tool to work with your build system may be the hardest part of the deployment; it is essential to ensure that this will work before spending money on a tool.

Commercial static analysis vendors generally offer a free trial, and you don't have to give the bugs back if you don't buy the tool. Another upside to testing on your real code is that if the tool is right for your organization, it will sell itself, by exhibiting bugs that convince even high-ranking sceptics that it is worth its price.

## IV. PITFALL 3: DON'T ACCEPT THE VENDOR'S DEFAULT CONFIGURATION

Once the static analysis vendor has configured your installation, investigate and experiment with its settings on your code. The defaults are likely to be a one-size-fits-all configuration, designed to make the tool look good and minimize problems for the vendor's support staff.

The vendor may have disabled some checkers because of a high rate of falsely reported defects ("false positives" in the jargon) in rare circumstances, or a lack of technical sophistication. Some of these checkers may nonetheless do well enough on your code to justify turning them on. Conversely, you may find that some checkers are irrelevant or unimpressive on your code. Some have statistical thresholds, which are probably set high to avoid false positives in mature open source code; the threshold for pre-release proprietary code may need to be set depressingly low. For example, if only a third of your developers check a function's return value, they may nonetheless be right; but the tool's default setting probably only reports a defect when a substantial majority perform this check.

I have had particularly good luck in enabling a careful selection of checkers in the tool's front end, which find problems similar to compiler warnings. Many of these might have been found by your compiler—but even with a strict warnings policy, many such bugs may slip through.

Sometimes warnings can be suppressed from the command line; sometimes an overly-strict policy requires that all warnings be reported, with the result that no-one looks through them for the genuinely serious ones.

## V. PITFALL 4: DON'T EXPECT AN IMMUTABLE CONFIGURATION

The experimentation recommended above requires considerable flexibility. Finding bugs with static analysis—like finding bugs in general—requires more adaptability, and less stability of process, than software development. If a configuration setting is discovered to be producing excessive false positives, or missing promising defects, it should be changed immediately. A change review process which makes sense for production code can stymie bug-finding, waste engineering time on false positives, and bring the tool into disrepute with those actually using it.

## VI. PITFALL 5: DON'T EXPECT STABLE METRICS

This flexibility will conflict with a natural management desire for stable metrics. If you disable a misguided checker, the number of reported defects will immediately go down; if you discover a new worthwhile checker, the number of defects will increase. This may conflict with management efforts to measure progress via the number of defects reported or processed. Dealing with this conflict requires a firm focus on the primary goal, of finding and fixing important bugs, and good judgment of quality; secondary considerations must remain secondary.

## VII. PITFALL 6: DON'T ACCEPT BROKEN ANALYSES

A static source code analysis tool (as opposed to byte-code or binary code analysis) emulates your compiler with its metacompiler, in order to transform your source code into a form that it can understand and find bugs in. A good front end to the metacompiler can do a remarkable job at this emulation, but will still not be perfect. Even small discrepancies in the emulation can lead to missing include files or rejected or misunderstood syntax, which in turn can lead to subtle but serious errors in analysis. The most prominent symptom is an excess of false positives, but missed bugs ("false negatives") are an even more serious, albeit invisible, symptom. Often a small number of underlying problems can lead to a majority of reported defects being false positives, and (even worse) a majority of potential defects being missed. Sometimes, on the other hand, the effects of an analysis problem will be less obvious, though still dangerous, with the misunderstood source code being skipped entirely.

Your tool may be configurable to stop and report failure for an analysis if a threshold of analysis errors is exceeded; if this threshold is adjustable, I recommend setting it quite low. If the tool cannot be prevented from continuing with a broken analysis, your build process should count the analysis errors in the tool's log files, and report failure if the number of errors per line surpasses a threshold. In either case, your first step when examining an analysis should be manually inspecting the tool's error logs. If there is no formal process to reject a broken analysis, a great deal of time can be wasted on flawed defects, and an unknown number of genuine bugs can be missed.

## VIII. PITFALL 7: DON'T JUST THROW THE TOOL OVER THE WALL

My first law of static analysis [Sheridan 2010a], [Yang 2009] is that developers who do static analysis voluntarily are the ones whose code needs it the least. Conversely, the developers whose code will benefit the most from static analysis will only use it if there is management backing, probably from the management champion discussed above. There are always excuses to deprioritize static analysis defects, some of them even valid [Pugh 2009]. Without serious management backing—in practice, not just in theory—for allocating time and personnel to investigating and fixing static analysis bugs, the tool is likely to remain shelf-ware.

## IX. PITFALL 8: DON'T PRESENT THE DEFECTS IN RANDOM ORDER

Most static analysis tools present defects in essentially random order (e.g., alphabetical by checker name, or by file). This avoids the need for judgment and minimizes problems for the vendor's support staff, but is a very poor presentational technique: You only get one chance to make a first impression, and if the first static analysis defect in a given engineer's queue is unimpressive, you may have lost him. [Kremenek & Engler 2003 p. 296.] This is particularly disappointing since there are techniques in the academic literature for ranking defects by reliability, relevance, and estimated importance. [Kremenek & Engler 2003], [Kremenek, Ashcraft, Yang, & Engler 2004]. It is particularly ironic that a co-author of these papers co-founded one of the leading static analysis vendors, which has rejected automated defect ranking. [Dawson Engler, personal communication, 11 May 2009].

Your developers will look at the tool's defects in some particular order; your setup and documentation should do as best you can to show important defects early, rather than likely false positives. Techniques for presenting defects vary widely; my approach is too long for presentation here [Sheridan 2010a]; the starting point was a central web page which was widely advertised, among other techniques, by a static analysis advertising poster depicting a large (literal) bug above the company pinball machine. More generally, a good static analysis tool is likely to find more bugs in your code than you have time and resources to fix, so good judgment in prioritization will be crucial. This too is outside the scope of this paper, but is one of the most important tasks when deploying static analysis. [Sheridan 2010b]

## X. PITFALL 9: DON'T EXPECT PERFECT ACCURACY

Any practical static analysis tool will produce false positives and, more important, don't-cares. Ensure that expectations are set low enough that developers continue to evaluate defects after they have encountered false positives. (On the other hand, do listen to complaints about such false positives, and improve the configuration to reduce them when practical.) Fixing bugs found via static analysis is likely to be far more efficient than conventional means, even with a very high false positive rate. Static analysis bypasses all of testers' time, all of an engineer's bug isolation/location time, and usually almost all of his or her analysis time: The tool goes right to the heart of the defect, and highlights the problem in clear, bright colors.

Fixing the problem, of course, still requires time and judgment, and not all static analysis bugs deserve to be fixed [Pugh 2009]. But much of the most time-consuming part of bug-hunting is bypassed. Even though rejecting false positives takes a small amount of time, and a disproportionate amount of developer resistance, do not allow imperfection to be an excuse to ignore bugs.

XI. PITFALL 10: DON'T ASSIGN YOUR MOST JUNIOR PROGRAMMERS TO EVALUATE STATIC ANALYSIS DEFECTS

It can be tempting to assign your least proficient programmers to classify and fix defects reported by static analysis. This is a mistake for two reasons. Evaluating static analysis defects is two levels of abstraction higher than writing code: Finding bugs in your code *should* be hard, and evaluating potential mistakes in such alleged bugs is harder still. Junior programmers may be all too ready to believe in habitual programming practices (or code which looks like it) rather than abstract objections to them. In particular, static analysis defects tend to occur on unexpected and confusing code paths, which programmers not accustomed to thinking abstractly will tend to reject uncomprehendingly.

Even more importantly, the most valuable benefit from static analysis, greater even than fixing bugs, is preventing future bugs, by educating the developer about his or her mistakes. If someone else is looking at these bugs, the developer never sees the mistakes and cannot learn from them.